\def\rn{}
\def\nn#1 #2{#2. #1}				
\def\nnn#1 #2 #3{#2. #3. #1}			
\def\nnnn#1 #2 #3 #4{#2. #3. #4. #1}		
\def\nnnnn#1 #2 #3 #4 #5{#2. #3. #4. #5. #1}	
\def\dualand{ and\hbox{ }}				
\def\multiand{ and,\hbox{ }}				
\def\rf#1;#2;#3;#4;#5 {{\frenchspacing\par\rn#1, #3 {\bf #4}, #5 (#2). \par}}
\def\rfbook#1;#2;#3;#4;#5 {{\frenchspacing\par\rn#1, {\it #3} (#5, #4, #2).\par}}
\def\rfprep#1;#2;#3 {{\par\frenchspacing\rn#1, #3 (#2).\par}}
\def\rfproc#1;#2;#3;#4;#5;#6 {{\frenchspacing\par\rn#1 #2, in {\it #3}, ed. #4 (#5: #6)\par}}
\def\rfprocp#1;#2;#3;#4;#5;#6;#7 {{\frenchspacing\par\rn#1 #2, in {\it #3}, ed. #4 (#5: #6), p#7\par}}
\def\dualand{ \&\hbox{ }}				
\def\multiand{ \&\hbox{ }}
				
\def\rf#1;#2;#3;#4;#5 {{\frenchspacing\par\rn#1, {\it #3} {\bf #4}, #5 (#2). \par}}

\def\etal{{\frenchspacing\it et al.}}
\def\ie{{\frenchspacing\it i.e.}}
\def\eg{{\frenchspacing\it e.g.}}
\def\etc{{\frenchspacing\it etc.}}

\def\beq#1{\begin{equation}\label{#1}}
\def\eeq{\end{equation}}
\def\beqa#1{\begin{eqnarray}\label{#1}}
\def\eeqa{\end{eqnarray}}

\def\fig#1{Fig.~#1}
\def\Fig#1{Fig.~#1}
\def\ZoomFig{1}
\def\rhoFig{2}
\def\kzFig{3}
\def\Pfig{4}

\def\spose#1{\hbox to 0pt{#1\hss}}
\def\simlt{\mathrel{\spose{\lower 3pt\hbox{$\mathchar"218$}}
     \raise 2.0pt\hbox{$\mathchar"13C$}}}
\def\simgt{\mathrel{\spose{\lower 3pt\hbox{$\mathchar"218$}}
     \raise 2.0pt\hbox{$\mathchar"13E$}}}
\def\simpropto{\mathrel{\spose{\lower 3pt\hbox{$\mathchar"218$}}
     \raise 2.0pt\hbox{$\propto$}}}

\def\ed{\end{document}}

\def\Ms{{\rm M}_\odot}

\documentstyle[prl,aps,epsf]{revtex}
\begin{document}
\twocolumn[\hsize\textwidth\columnwidth\hsize\csname@twocolumnfalse\endcsname


\preprint{IASSNS-AST 97/666}

\title{Measuring Spacetime: from Big Bang to Black Holes}

\author{Max Tegmark}

\address{Dept. of Physics, Univ. of Pennsylvania, 
Philadelphia, PA 19104; max@physics.upenn.edu}

\date{A slightly abbreviated version of this paper was published as an invited review
in {\it Science}, {\bf 296}, 1427-1433 (2002)
}

\maketitle

\begin{abstract}
Space is not a boring static stage on which events unfold over
time, but a dynamic entity with curvature, fluctuations and a rich life of its own
which is a booming area of study.
Spectacular new measurements of the cosmic microwave background,
gravitational lensing, 
type Ia supernovae,
large-scale structure, spectra of the
Lyman $\alpha$ forest,
stellar dynamics and
x-ray binaries are probing the properties of spacetime over 22 orders of magnitude
in scale.
Current measurements are consistent with an infinite flat everlasting universe containing
about 30\% cold dark matter, 65\% dark energy 
and at least two distinct populations of black holes.
\end{abstract}


] 


\section{Introduction}

Traditionally, space was merely a three-dimensional static stage where the cosmic drama played out over time.
Einstein's theory of general relativity \cite{Einstein,WeinbergBook,MTW} replaced this by four-dimensional spacetime,
a dynamic geometric entity with a life of its own,
capable of expanding, fluctuating and even curving into black holes.
Now the focus of research is increasingly shifting from
the cosmic actors to the stage itself. 
Triggered by progress in detector, space and computer technology,
an avalanche of astronomical data is revolutionizing our ability to 
measure the spacetime we inhabit on scales ranging from the
cosmic horizon down to the event horizons of suspected black holes,
using photons and astronomical objects as test particles.
The goal of this article is to review these measurements and future prospects, focusing
on four key issues:
\begin{enumerate}
\item The global topology and curvature of space
\item The expansion history of spacetime and evidence for dark energy
\item The fluctuation history of spacetime and evidence for dark matter
\item Strongly curved spacetime and evidence for black holes
\end{enumerate}  
In the process, I will combine constraints from
the cosmic microwave background \cite{CMB},
gravitational lensing, 
supernovae Ia,
large-scale structure, the
Lyman $\alpha$ forest\cite{LyAF},
stellar dynamics and
x-ray binaries.
Although it is fashionable to use cosmological data to measure a small number
of free ``cosmological parameters'', I will argue that improved data allow raising the
ambition level beyond this, testing rather than assuming
the underlying physics. 
I will discuss how with a minimum of assumptions, 
one can measure key properties of spacetime itself in terms of 
a few cosmological functions:
the expansion history of the universe, the spacetime fluctuation spectrum and its growth.


\begin{figure}[pbt]
{\epsfxsize=9.0cm\epsffile{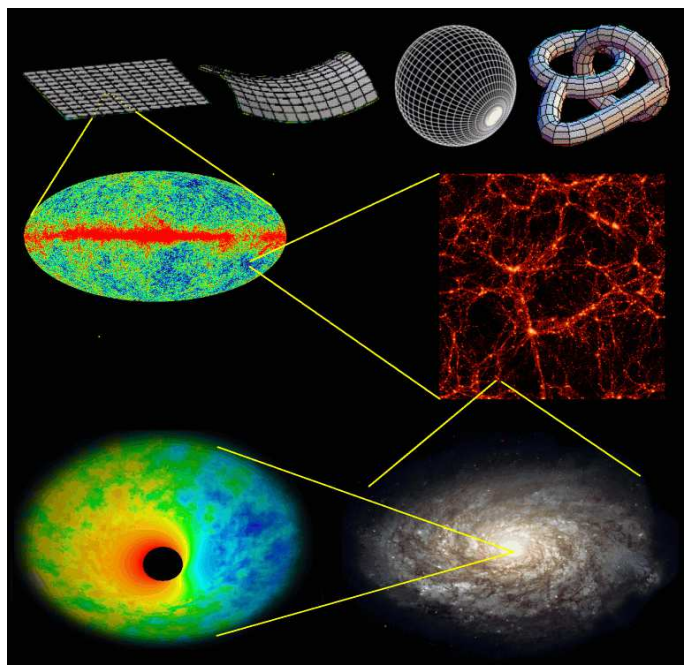}}
\smallskip
\caption{
Summary of the spacetime issues discussed in this article.
One can use photons and astronomical objects as test particles to measure spacetime over 
22 orders of magnitude in scale,
ranging from the cosmic horizon
(probing the global topology of and curvature of space --- top)
down to galaxies (giving evidence for dark matter), 
galactic nuclei and binary stellar systems (giving evidence for black holes).
The figure illustrates how spacetime ripples at the $10^{-5}$ level will be imaged by the
cosmic microwave background satellite MAP \protect\cite{MAP} and has
grown via gravitational instability into cosmic large-scale structure \protect\cite{VIRGO},
galaxies and, it seems, black holes \protect\cite{Bromley}.
}
\label{zoomFig}
\end{figure}

\subsection{Goals and tools}

Before embarking on our survey of spacetime, let us briefly review what it is we want to measure,
the basic tools at our disposal \cite{WeinbergBook,MTW,WillBook} and the broad-brush picture
of how our topics fit together.
According to general relativity, spacetime is what mathematicians call a manifold, 
characterized by a topology and a metric.
The topology gives the global structure (\fig{\ZoomFig}, top): 
is space infinite in all directions 
like in high-school geometry
or multiply 
connected like say a hypersphere or doughnut so that traveling in a straight line could in principle
bring you back home --- from the other direction?
The metric determines the local shape of spacetime, \ie,  the distances and time intervals we 
measure, and is mathematically specified by a $4\times 4$ matrix at each 
point in spacetime.

General relativity theory (GR) consists of two parts, each providing a tool for measuring the metric.
The first part of GR states that in the absence of non-gravitational forces,
test particles (objects not heavy enough to have a noticeable effect on the metric)
move along geodesics in spacetime, generalized straight lines, so the observed motions of
photons and astronomical objects allow the metric to be reconstructed. 
I will refer to this as {\it geometric} measurements of the metric.
The second part of GR states that the curvature of spacetime (expressions involving
the metric and its first two derivatives) is related to its matter content
--- in most cosmological situations simply the density and pressure, but 
sometimes also bulk motions and stress energy.
I will refer to such measurements of the metric as  {\it indirect}, because they 
assume the validity of the Einstein field equations of GR.

\subsection{The broad brush picture}

The current consensus in the cosmological community is that spacetime is 
extremely smooth, homogeneous and isotropic (translationally and rotationally invariant)
on large ($\sim 10^{23}$m$-10^{26} $m) scales, with small fluctuations that have grown 
over time to form objects like galaxies and stars on smaller scales.
Cosmic Microwave Background (CMB) observations have shown \cite{Smoot92}
that space is almost isotropic on the scale of our cosmic horizon ($\sim 10^{26} $m),
with the metric fluctuating by only about 
one part in $10^5$ from one direction to another, and 
combining this with the so-called cosmological principle, the assumption
that there is nothing special about our vantage point, implies that 
space is homogeneous as well.  Three-dimensional maps 
of the galaxy and quasar distribution give more
direct evidence for large-scale homogeneity \cite{Colless01,Zehavi01,Hoyle01}.

The fact that the CMB fluctuations are so small is useful,
because it allows the intimidating nonlinear partial differential equations governing spacetime 
and its matter content to be accurately solved using a perturbation expansion.
To zeroth order (ignoring the fluctuations), this fixes the global metric to be of the so-called
Friedman-Robertson-Walker (FRW) 
form, which is completely 
specified except for a curvature parameter and a free function giving its expansion history.
To first order, density perturbations grow due to gravitational 
instability and gravitational waves 
propagate through the FRW background spacetime, all governed by 
{\it linear} equations.
Only on smaller scales ($\simlt 10^{23}$m) do the fluctuations
get large enough that nonlinear dynamics becomes important --- in 
the realm of galaxies, stars and, perhaps, black holes.
This review is organized analogously: Sections 2 and 3 discuss spacetime to 
0th order (curvature, topology and expansion history), Section 4 describes
spacetime to 1st order (fluctuations) and Section 5 focuses on nonlinear objects,
mainly black holes.

\section{Overall shape of spacetime}

\subsection{Curvature of space}
 
The question of whether space is infinite was answered last year with a resounding {\it maybe}.
For an FRW metric, answering this question is equivalent to measuring the 
curvature of space as illustrated by the top left three cases in \fig{\ZoomFig}, specifically 
a single number $R$ known as the {\it radius of curvature}. 
$R$ is the radius of the hypersphere if space is finite, 
$R=\infty$ if space is flat, and 
$R$ is an imaginary number($R^2<0$) for saddle-like curvature.
Because the three angles of a triangle will add up to $180^\circ$ in flat space,
more if $R^2>0$ (like on a sphere) and less if $R^2<0$ (like on a saddle)
cosmologists have measured $R$ using the largest triangle 
available: one with us at one corner and the other two corners
on the hot opaque surface of ionized hydrogen that delimits the visible universe
and emits the CMB, merely 400,000 years after the big bang. 
Photographs of this surface reveal hot and cold spots of a characteristic angular size that
can be predicted theoretically. This characteristic spot size  
(or, more rigorously, the first peak in the CMB power spectrum \cite{powerspectrum}) 
subtends about $0.5^{\circ}$ --- like the Moon  --- if space is flat.
Sphere-like curvature would make all angles appear larger, so 
characteristic spots much larger than the Moon would indicate a finite universe 
curving back on itself, whereas smaller spots would indicate infinite
space with negative curvature.

By 1994, evidence was mounting that there really was a peak in
the CMB power spectrum \cite{ScottSilkWhite95}, or at least a rise
towards smaller scales.
Data kept improving, and in 1998 the Toco experiment provided the 
first unambiguous detection and localization of a peak. 
The BOOMERanG experiment measured it with great precision in 2000, and 
by now the BOOMERanG, DASI and MAXIMA \cite{acronyms} 
teams have all seen both this peak
and hints of additional smaller scale peaks matching theoretical predictions.

So is the universe infinite? The answer so far is still maybe, because the 
characteristic spot size has turned out to be 
so close to $0.5^\circ$ that we still  cannot tell whether space is 
perfectly flat or very slightly curved either way. 
The sharpest current limits on the curvature radius, obtained by 
combining all CMB experiments with galaxy clustering data \cite{consistent,Efstathiou02} 
to constrain other parameters affecting the spot sizes (mainly the cosmic matter budget),
are $|R| > 20 h^{-1}$Gpc$\,\approx 10^{27}$m.
This is in sharp contrast to a few years ago, when the 
most popular models had negatively curved space with 
$|R|\approx  4 h^{-1}$ Gpc.
In other words, space now seems to be either infinite or much larger than the observable universe,
whose radius is about $9 h^{-1}$ Gpc.

In 1900, Karl Schwarzschild discussed the possibility that space was curved
and published a paper with a lower limit $R>2500$ light-years
$\approx  2\times 10^{19}$m \cite{Schwarzschild}.
A century later, we thus know that the universe is at least
another 40 million times larger!

\subsection{Topology of space}

Even if space turns out to be negatively curved or perfectly flat,
it might be finite. 
General relativity does not prescribe the global topology, so 
various possibilities are possible (\fig{\ZoomFig}, top).
The simplest non-trivial model has flat space and the topology of a 
three-dimensional torus, where opposing faces of a 
cube of size $L\times L\times L$ are identified to be one and the same.
Living in such a universe would be indistinguishable from living in a perfectly periodic one:
if $L=10$\,m, you could see the back of your own head 10 meters away, and additional copies
at 20\,m, 30\,m, and so on --- searches for multiple images of cosmological objects
have constrained such models \cite{Roukema}.
Also, just as a finite guitar string has a fundamental tone and overtones, 
linear spacetime fluctuations in 
such a toroidal universe could have only certain discrete wavenumbers. 
As a result, its CMB power spectrum would differ on large scales, and
the COBE\cite{acronyms} data was used to show that if the
universe were such a torus, then $L$ must be at least of 
the order of the cosmic 
horizon \cite{Stevens93,BBtopo1}. 
Indeed, it was shown that all three dimensions of the torus must at least about this large 
to explain the absence of a type of approximate reflection symmetry in the COBE map \cite{BBtopo2}.
This early work triggered dozens of papers in so-called cosmic crystallography, which turned out to 
be a rich mathematical subject --- for an up-to-date review, see \cite{Levin01}.
For instance, circles in the sky with near-identical temperature patterns were shown to be smoking-gun
signals of compact topology.
Unfortunately from an aesthetic point of view, many of the most mathematically elegant models,
negatively curved yet compact spaces, have been abandoned after the recent evidence 
for spatial flatness. NASA's Microwave Anisotropy Probe(MAP) \cite{MAP}
will allow the cosmic topology to be probed with
a new level of precision. 

The interim conclusion about the overall shape of space is thus ``back to basics'' :
although mathematicians have discovered a wealth of complicated manifolds
to choose from and both positive and negative curvature would have been allowed {\it a priori},
all available data so far is consistent with the simplest possible space, 
the infinite flat Euclidean space that we learned about in high school.
That is in regards to three-dimensional space. The global structure of our four-dimensional 
spacetime also depends on the beginning and end of time, to which we turn in the next section.

\begin{figure}[pbt]
\centerline{\epsfxsize=9.0cm\epsffile{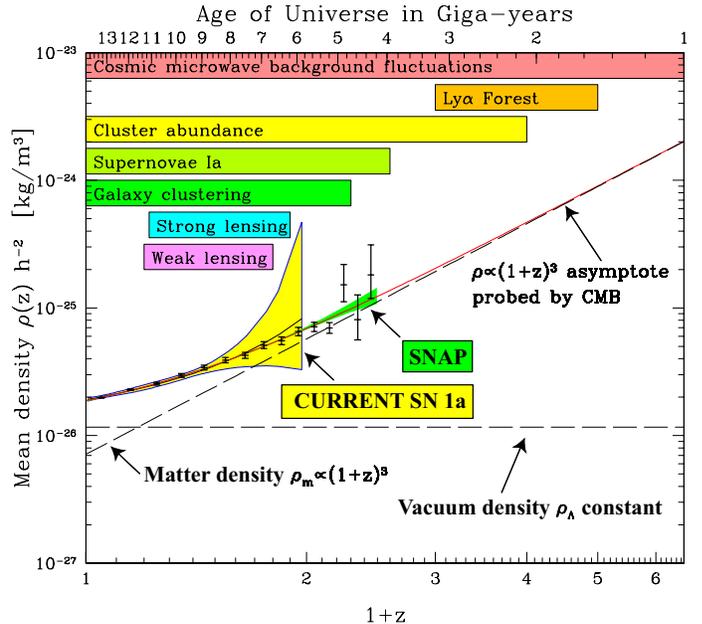}}
\caption{
Solid curve shows the concordance model \protect\cite{consistent}
for the evolution of the cosmic mean density $\rho(z)\propto H(z)^2$.
This curve uniquely characterizes the spacetime expansion history.
The horizontal bars indicate the rough
redshift ranges over which the various cosmological probes discussed
are expected to constrain this function.
Because the redshift scalings of all density contributions except that of
dark energy are believed to be straight lines with known slopes
in this plot (power laws),
combining into a simple quartic polynomial,
an estimate of the
dark energy density $\rho_X(z)$ can be readily extracted from 
this curve. 
Specifically, 
$\rho\propto (1+z)^4$ for the cosmic microwave background (CMB), 
$\rho\propto (1+z)^3$ for baryons and cold dark matter,
$\rho\propto (1+z)^2$ for spatial curvature,
$\rho\propto (1+z)^0$ for a cosmological constant
and $\rho\propto (1+z)^{3(1+w)}$ for dark energy with a constant equation
of state $w$. Measurement errors are for current SN Ia constraints (yellow band) and 
a forecast for what the SNAP satellite \protect\cite{SNAP} can do (green band),
assuming flat space as favored by the CMB.
Error bars are for a non-parametric reconstruction with SNAP.
}
\label{rhoFig}
\end{figure}

\section{Spacetime expansion history}

One of the key quantities that cosmologists yearn to measure is the function $a(t)$, describing
the expansion of the universe over time 
--- if space is curved, $a$ is simply the magnitude of the radius of curvature, $a=|R|$.
A mathematically equivalent function more closely 
related to observations is the Hubble parameter as a function of redshift, $H(z)$,
giving the cosmic expansion rate and defined by 
$H\equiv {d\over dt}\ln a$, $1+z\equiv a(t_{\rm now})/a(t)$.
Let us first discuss how this function encodes key information about the cosmic matter budget, the origin and the ultimate fate of the universe, and then turn to how it 
can be, has been and will be measured.

\subsection{What $\rho(z)$ tells us about dark energy}

As illustrated in Figure 2, 
squaring our curve $H(z)$ gives us the cosmic matter density.
If the Einstein Field equations of GR are correct, then 
the mean density of the universe is given by the Friedmann equation \cite{WeinbergBook}
\beq{FriedmannEq}
\rho(z)= {3H(z)^2\over 8\pi G}.
\eeq
Here $G$ is Newton's gravitational constant and, if space is curved, 
$\rho$ is defined to include an optional
curvature contribution $\rho_{\rm curv}\equiv - {3c^2\over 8\pi G R^2}$,
where $c$ is the speed of light.
Conveniently, all standard components of the cosmic matter budget 
contribute simple straight lines to this plot, because their 
densities drop as various power laws as the universe expands.
For instance, the densities of both ordinary and cold dark matter particles
are inversely proportional to the volume of space, scaling as 
$\rho\propto (1+z)^3$. 

\Fig{\rhoFig} shows that although the cosmic density $\rho(z)$ measured from 
SN Ia and CMB was indeed higher in the past, the curve rises slower than this towards higher redshift, with a shallower 
slope than $3$ at recent times.
This is evidence for the existence of {\it dark energy}, a substance whose density does not rise rapidly with $z$. Adding a cosmological 
constant contribution 
$\rho_\Lambda\approx 4\times 10^{-26}$ kg/m$^3$ 
(about $2/3$ of the current matter budget) whose density is, by definition, constant, provides 
a good fit to the measurements (\fig{\rhoFig} ). 
This discovery, made independently by two teams in 1998 \cite{Perlmutter98,Riess98}, stunned the scientific community and triggered a worldwide effort to determine the nature
of the dark energy. 
A model-independent approach will be to 
measure the curve $\rho(z)$ more accurately with a variety of different techniques as
illustrated in the figure and described below, thereby answering two separate questions;
\begin{enumerate}
\item Do independent measurements of $\rho(z)$ agree, so that we can rule out problems with
observations and their interpretation?
\item Subtracting out the slope 3 line contributed by ordinary and dark matter, 
what is the time-dependence of dark energy density $\rho_X(z)$? 
If it is constant, we may have measured vacuum energy/Einstein's cosmological constant,
and if not, we should learn interesting physics about  a new scalar quintessence field, 
or whatever is responsible.
\end{enumerate}
A less ambitious approach that is currently popular is assuming that 
the equation of state (pressure-to-density ratio) $w$ of the dark energy is constant
\cite{Garnavich98,Maor00,Huterer00},
which is equivalent to assuming that $\rho_X(z)$ is a straight line in
\fig{\rhoFig} with a free amplitude and slope.

\subsection{What $\rho(z)$ tells us about our origin and destiny}

If we can understand the different components of the cosmic matter budget well enough to extrapolate
the curve $\rho(z)$ 
from \fig{\rhoFig}
to the distant past and future, 
we can use the Friedmann equation to solve for $a(t)$ and 
obtain information about the origin and ultimate fate of spacetime. 
$a(t)=0$ 
in the past or future would correspond to a singular big bang or big crunch, 
respectively, with infinite density $\rho(z)$. 
As to the past, such extrapolation seems justified at least back to the first seconds after the
big bang, given the success of big bang nucleosynthesis in accounting for the primordial 
light element abundances \cite{Burles01,Carroll02}. 
Regarding the very beginning, the jury is still out. 
Extrapolation back to the very beginning is more speculative. 
According to the currently most popular scenario, 
a large and nearly constant value of $\rho$ at   
$t\simlt 10^{-34}$ seconds caused exponential expansion 
$a(t)\propto e^{Ht}$ during a period known as inflation \cite{GuthSteinhardt84,LindeBook,LiddleBook}, successfully 
predicting both negligible spatial curvature and, as discussed in the next section, 
a nearly scale-invariant adiabatic scalar power spectrum\cite{powerspectrum}
with subdominant gravitational waves. 
A rival 
``ekpyrotic'' 
model inspired by string theory and a related eternally oscillating model have 
attracted recent attention\cite{Khoury,LindeKallosh,SteinhardtTurok}.
If the density approaches the Planck density ($10^{97}\>$kg$/$m$^3$)
as $t\to 0$, quantum gravity effects for which 
we lack a fundamental theory should be important, and a host of speculative scenarios have been
put forward for what happened at $t\sim 10^{-43}$ seconds. A very incomplete sample includes
the Hawking-Hartle no-boundary condition \cite{Hawking}, 
God creating the universe, 
the universe creating itself \cite{Gott}, and so-called 
pre-big-bang models \cite{Venetiano}.
Another possibility is that the Planck density was never attained and that there was no beginning, 
just an eternal fractal mess of replicating inflating bubbles, with our observed spacetime being 
merely one in an infinite ensemble of regions where inflation has stopped \cite{LindeBook,Vilenkin}.

As to the distant future, the expansion can clearly only stop ($H=0$) if the effective
density $\rho(z)$ drops to zero. The only two density contributions that can in principle be negative
are those of curvature (which now seems to be negligible) and dark energy (which seems to be positive),
suggesting that the universe will keep expanding forever. Indeed, 
if the dark energy density stays constant, 
we are now entering another inflationary phase of exponential expansion 
($a(t)\propto e^{Ht}$), and in about $10^{11}$ years, 
our observable universe will be dark and lonely with almost all extragalactic 
objects having disappeared across our cosmic horizon \cite{Loeb02}. 
However, such conclusions must clearly be 
taken with a grain of salt 
until the nature of dark energy is understood. 

\subsection{How to measure $\rho(z)$}

In conjunction with the curvature radius $R$, 
the curve $H(z)$ can be measured purely geometrically, using photons as test
particles. Given objects of known luminosity (``standard candles'') or known 
physical size (``standard yardsticks'') at various redshifts, one simply
compares their measured brightness or angular size with theoretical
predictions. Because predictions, which follow from
computing the trajectories of nearly parallel light rays,
depend on only $H(z)$ and the (apparently negligible) curvature of space,
objects at multiple redshifts can be used to
reconstruct the curve $H(z)$ \cite{WangGarnavich01,gravity}.

The best standard candles to date are supernovae of type Ia, and the
92 SN Ia published by the two search teams \cite{Perlmutter98,Riess98} were used \cite{gravity}
to measure $H(z)$ and thereby $\rho(z)$ in \fig{\rhoFig}.
These cosmic bombs all have the same mass, since they result when
a white dwarf accretes enough gas from a companion star to exceed the
Chandrashekar mass limit of 1.4 solar masses. 
They therefore have similar
luminosity, and it has been shown that their actual luminosity can be accurately 
calibrated using their dimming rate and color \cite{Perlmutter98,Riess98}.
The best standard yardstick so far is the characteristic CMB spot size discussed above,
suggesting that space is flat. As reviewed in \cite{RowanRobinsonBook,gravity}, 
numerous other candles and yardsticks have been discussed,
especially in the Hubble parameter literature \cite{Freedman01} focused on measuring
$H(z)$ for $z\approx 0$,
and although many have proven hard to standardize because of issues like galaxy evolution,
it is far from clear that new multicolor surveys will not be able to 
measure $H(z)$ independently of SN Ia.

$H(z)$ can also be measured indirectly. 
As discussed in the next section, $H(z)$ affects the growth of density fluctuations
and can therefore be probed by galaxy clustering and other techniques as indicated in \fig{\rhoFig}.
Such fluctuation measures have constrained matter 
to make up no more than about a third of the critical density needed to explain why space is
flat. This Enron-like accounting situation 
provides supernova-independent evidence for dark energy 
\cite{consistent,Efstathiou02,CosmicTriangle}.


\section{Growth of cosmic structure}


While SN Ia and CMB peak locations
have recently revolutionized our knowledge of the metric
to 0th order (curvature, topology and expansion history), other observations 
are probing its 1st order fluctuations with unprecedented accuracy.
These perturbations come in two important types. The first are gravitational waves,
hitherto undetected ripples in spacetime that propagate at the speed 
of light without growing in amplitude.
The second are density fluctuations, which can get amplified by gravitational instability
(\fig{\ZoomFig})
and are being measured by 
CMB, gravitational lensing and the clustering of extragalactic objects, 
notably galaxies and gas clouds absorbing quasar light (the so-called 
Lyman $\alpha$ forest, Ly$\alpha$F) over a range of 
scales and redshifts (\fig{\kzFig}). 

\begin{figure}[phbt]
\centerline{{\vbox{\epsfxsize=8.6cm\epsfbox{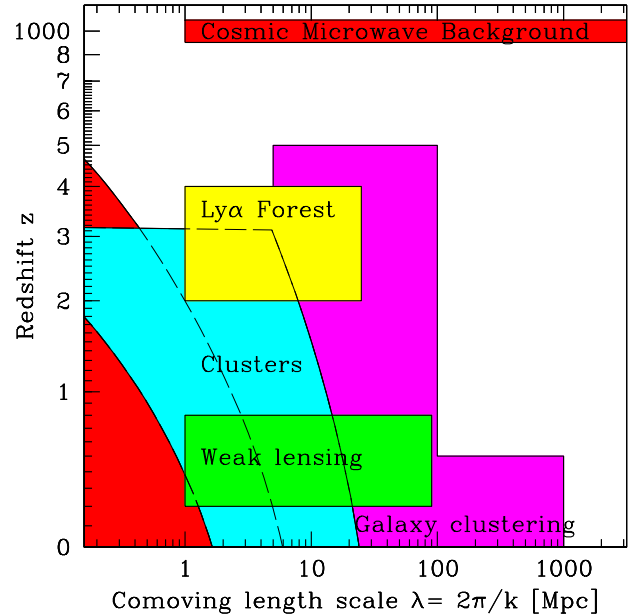}}}}
\smallskip\smallskip
\caption{
Shaded regions show ranges of scale and redshift over which
various observations are likely to measure spacetime fluctuations 
over the next few years.
The lower left region, delimited by the dashed line, is the non-linear
regime where rms density fluctuations exceed unity in
the  ``concordance'' model from 
\protect\cite{consistent}.
}
\label{kzFig}
\end{figure}

Plane wave perturbations of different wavenumber 
evolve independently by linearity, and are so far 
consistent with having uncorrelated 
Gaussian-distributed amplitudes 
\cite{MaximaNonGauss} as predicted by 
most inflation models \cite{LiddleBook}.
The 1st order density perturbations are therefore characterized by a single function
$P(k,z)$, the {\it power spectrum} \cite{powerspectrum}, which gives the variance of the fluctuations 
as a function of wavenumber $k$ and redshift $z$.
$P(k,z)$ depends on (and can therefore teach us about) three things:
\begin{enumerate}
\item The cosmic matter budget 
\item The seed fluctuations created in the Early Universe
\item Galaxy formation: reionization, ``bias'', {\etc}
\end{enumerate}
A key challenge is to robustly disentangle the three. We are not there yet, but 
new data is making this increasingly feasible 
because each of the probes 
in \fig{\kzFig} 
involve different physics and is affected by the three
in different ways 
as outlined below.

Given the profusion of recent measurements of $H(z)$ and $P(k,z)$, it is 
striking that there is a fairly simple model that currently seems to fit everything
(\fig{\rhoFig} and \fig{\Pfig}).
In this so-called concordance model \cite{consistent,Efstathiou02,CosmicTriangle,Lahav01},
the cosmic matter budget consists of about 
5\% ordinary matter (baryons), 30\% cold dark matter, 0.1\% hot dark matter (neutrinos)
and 65\% dark energy based on CMB and LSS observations, in good agreement with
Ly$\alpha$F \cite{consistent,Croft00,Hannestad01}, lensing \cite{WeakLensingReview,Wittman00,Waerbeke00,Bacon00,Kaiser00,Rhodes01,Waerbeke01}
and SN Ia \cite{Perlmutter98,Riess98}. 
The seed fluctuations created in the early universe are consistent with the inflation prediction of a
simple power law $P(k,z)\propto k^n$ early on, with $n=0.9\pm 0.1$ \cite{consistent,Efstathiou02}.
Galaxy formation appears to have heated and reionized the universe 
not too long before redshift $z=6$ based on the Ly$\alpha$F \cite{Becker01,Djorgovski01}.

\begin{figure}[pbt]
\vskip-0.6cm
\centerline{\epsfxsize=9.0cm\epsffile{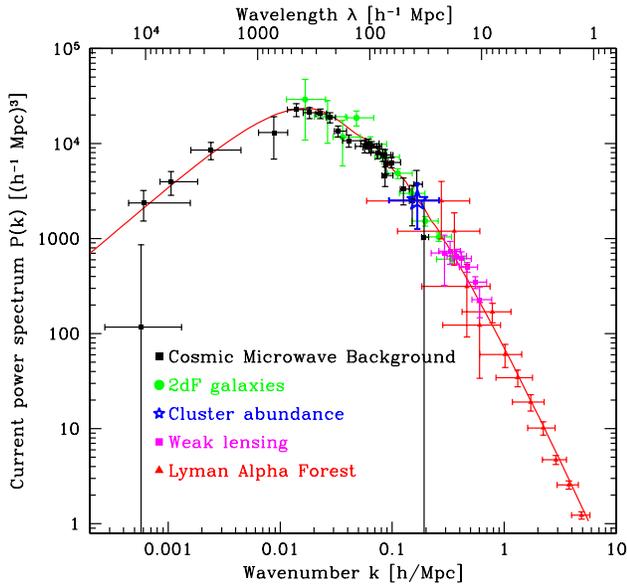}}
\vskip-0.5cm
\caption{
Measurements of the current $(z=0)$ power spectrum of density fluctuations
computed as described in \protect\cite{pwindows}, assuming the matter budget
of \protect\cite{Efstathiou02} and reionization at $z=8$.
The CMB measurements combine the information from all experiments to date as 
in \protect\cite{pwindows}.
LSS points are from a recent analysis \protect\cite{2df} of
the 3D distribution of 2dF galaxies \protect\cite{Colless01}, 
and correcting them for bias shifts them 
vertically ($b=1.3$ assumed here)
and should perhaps blue-tilt them slightly.
The cluster error bars reflect the spread in the literature.
The lensing points are based on \protect\cite{Hoekstra02}.
The Ly$\alpha$F points are from a reanalysis
\protect\cite{GnedinHamilton01} of \protect\cite{Croft00} and have an overall calibration
uncertainty around 17\%. 
The curve shows the concordance model of \protect\cite{Efstathiou02}.
}
\label{zrhoFig}
\end{figure}

Although the mere existence of a concordance model is a striking success,
inferences about 
things like 
the expansion history, the matter budget and the early universe
involve
many assumptions --- about the nature of dark energy and dark matter
(\eg, interactions, temperature, pressure, sound speed, viscosity \cite{HuGDM}),
about gravity, about galaxy formation, and so on.
Since the avalanche of new cosmology data is showing no
sign of slowing down, it is becoming feasible to to raise the ambition level to 
test rather than assume the underlying physics, probing the 
nature of dark energy, dark matter and gravity.
Given the matter budget and the expansion history $H(z)$, theory 
predicts the complete time-evolution of linear clustering, so 
measuring its redshift dependence (\fig{\kzFig}) offers redundancy
and powerful cross-checks.

Let us briefly summarize the status of our five power spectrum probes in \fig{\kzFig}.
{\bf Gravitational lensing} 
uses photons from distant galaxies as test particles to 
measure the metric fluctuations caused by intervening matter, 
as manifested by distorted images of distant objects.
The first measurements of $P(k,z)$ with this ``weak lensing'' technique \cite{WeakLensingReview}
were reported in 2000 
\cite{Wittman00,Waerbeke00,Bacon00,Kaiser00,Rhodes01,Waerbeke01}.
%
3D mapping of the universe with 
{\bf galaxy redshift surveys} 
offers another window on
the cosmic matter distribution, through its gravitational effects on galaxy clustering.
This field is currently being transformed by the 2 degree Field (2dF) survey and the
Sloan Digital Sky Survey, which will jointly map 
more than $10^6$ galaxies, 
and complementary surveys 
will map high redshifts and the evolution of clustering.
Additional information can be extracted from galaxy velocities \cite{Zehavi99}.
The abundance of 
{\bf galaxy clusters} 
at different epochs, as probed by 
optical, x-ray, CMB or gravitational lensing surveys,
is a sensitive probe of $P(k,z)$ on smaller scales \cite{Bahcall98,Pierpaoli00,Holder01}
and the 
{\bf Ly$\alpha$F} 
offers a new and exciting probe of matter clustering on still smaller scales
when the universe was merely 10-20\% of its present age
\cite{Croft00,McDonald00,lya,GnedinHamilton01}. 
%
{\bf CMB} 
experiments probe $P(k,z)$ through 
a variety of effects as far back as to redshifts $z>10^3$
\cite{HuNature97,cmbfast}.
The MAP satellite will publicly release CMB temperature measurements of unprecedented
quality in December 2002 \cite{MAP}, and two 
new promising CMB fronts are opening up --- CMB polarization (still undetected) 
and CMB fluctuations on tiny (arcminute) angular scales.

There is a rich literature on how all these complementary probes can be combined to
break each others' degeneracies and independently measure 
the matter budget, the primordial power spectrum and galaxy formation details 
\cite{WangGarnavich01,gravity,HuGDM,parameters2,gdm}, so I will merely give a few examples here.
The power spectra measured by CMB, LSS, lensing and Ly$\alpha$F are the product
of the three terms: (i) the primordial power spectrum, (ii) a 
so-called transfer function
quantifying the
subsequent fluctuation growth, and (iii) (for LSS and Ly$\alpha$F only) a 
so called bias factor
accounting for the fact that the measured galaxies/gas clouds may cluster differently than
the underlying matter.

{\bf Disentangling bias and systematic errors:}
Galaxy bias has now been directly measured from data and found to be of order unity
for typical 2dF galaxies \cite{Lahav01,Verde01}, and Ly$\alpha$F bias may be
computable with hydrodynamics simulations \cite{Croft00,lya,GnedinHamilton01}.
Although CMB, LSS, lensing and Ly$\alpha$F each comes with 
caveats of their own, their substantial overlap (\fig{\kzFig}) should allow 
disagreements between data sets to be distinguished from 
disagreements between data and theory.

{\bf Disentangling primordial power from the matter budget:}
The transfer function can be disentangled from the primordial power 
because it depends on the matter budget, and conveniently in rather opposite
ways for CMB than for low redshift $P(k)$ measurements (LSS, lensing, Ly$\alpha$F).
For instance, increasing the cold dark matter density $h^2\Omega_c$ shifts the galaxy power
spectrum up to the right and the CMB peaks down to the left if the 
primordial spectrum is held fixed.
Adding more baryons boosts the odd-numbered CMB peaks
but suppresses the galaxy power spectrum rightward of its peak and also makes it wigglier. 
Increasing the dark matter percentage that is hot (neutrinos)
suppresses small-scale galaxy power while leaving the CMB almost unchanged.
This means that combining CMB with other data allows unambiguous determination
of the matter budget, and the primordial power spectrum can then be inferred.
Combining CMB temperature and polarization measurements also helps in this
regard, because the characteristic wiggles imprinted by the baryons and dark matter are
out of phase for the two, whereas wiggles due to the primordial spectrum would of course
line up for the two \cite{pwindows}.

Although the best is still to come in this area, the basic conclusion that the universe is awash in 
nonbaryonic dark matter already appears quite solid, supported independently by
CMB, Ly$\alpha$F, galaxy surveys, cluster counting and lensing --- and by additional 
evidence in the next section. The agreement on the baryon density between 
fluctuation studies (CMB + galaxy surveys) and nucleosynthesis
and on the dark energy density between fluctuation studies and SN Ia
are both indications that spacetime fluctuation measurements are on the right track
and will live up to their promise in this decade of precision cosmology.

\section{Nonlinear clustering \& black holes}

On small scales, the linear perturbation expansion eventually breaks down
as density fluctuations grow to be of order unity, collapsing to form 
a variety of interesting astrophysical objects. Although the theoretical predictions 
are more difficult in this regime, the metric can still be accurately measured using 
photons and astrophysical objects as test particles.
The gravitational potential well is probed by strong gravitational lensing of photons through its
distorting effect on background objects \cite{StrongLensingReview}
and also by the motions of massive objects like galaxies, stars or gas clouds.
The orbital parameters in a binary system reveal the masses of the two objects,
just as we once weighed the Sun by exploiting Earth's orbit around it.
In more complicated systems, the central mass distribution can be inferred 
statistically from velocity dispersions observed in the vicinity.
Below I review how these basic tools have revealed 
surprises on three 
vastly 
different scales: 
dark matter in galaxies and clusters ($\sim 10^{20-23}$m),
supermassive black holes in galactic bulges ($\sim 10^{10}-10^{13}$m) 
and stellar-mass black holes ($\sim 10^4-10^5$m).
Recent black hole reviews include 
\cite{Celotti99,KrolikBook,Kormendy01,Marconi02,KingBook}.

\subsection{Dark matter in galaxies and clusters}

As noted by Zwicky in 1933 \cite{Zwicky33}, 
the amount of mass in galaxies and galaxy clusters inferred from
rotation curves or velocity dispersions exceeds the mass of 
luminous matter by a large factor. 
Precision measurements with a variety of techniques have confirmed 
this finding, providing evidence that both galaxies and clusters are accompanied by
roughly spherical halos of cold dark matter. This dark matter evidence is 
independent of that from linear perturbation theory described above, yet produces 
roughly consistent estimates of the total cosmic dark matter density
\cite{Hoekstra01,BahcallML}.

New measurements such as mapping tidal streamers, stripy remnants of galaxies 
cannibalized by the Milky Way in the past,
are raising the ambition level towards a full 3D 
reconstruction of our own dark matter halo,
and early results suggest that it may be elliptical rather than perfectly spherical \cite{SDSSstreamers}.
Measurements of the shape and substructure of dark matter halos can probe
the detailed nature of the dark matter.
Indeed, computer simulations with cold dark matter composed of weakly interacting 
particles appear to predict overly dense cores in the centers of galaxies and clusters,
and that there should be about $10^3$ discrete dark matter halos in our
Galactic neighborhood (the Local Group), in contrast to the less than $10^2$ galaxies actually observed. 
These halo profile and substructure problems have triggered talk of a 
cold dark matter crisis and much recent interest in 
self-interacting dark matter \cite{Spergel00},
warm dark matter \cite{Bode00}
and other more complicated dark matter models which suppress cores and substructure.
It is not obvious that there really is a crisis, since baryonic feedback properties 
may be able to reconcile vanilla cold dark matter with observations and since
substantial halo substructure has recently been detected with gravitational lensing \cite{Dalal02},
but this active research area should teach us more about dark matter properties whatever 
they turn out to be.

\subsection{Supermassive black holes}

Karl Schwarzschild was allegedly so distressed by his 1916 
solution to the Einstein field equations that he hoped that such 
sinister objects, later christened black holes by Wheeler, did not exist in 
the real universe. The irony is that
monstrous black holes 
are nowadays considered the {\it least} exotic explanation for 
the phenomena found in the centers of most --- if not all --- massive galaxies.

The spatial and velocity distribution of stars have unambiguously revealed compact objects
weighing $10^6-10^{10}$ solar masses at the centers of over a dozen galaxies.
The most accurate measurements are for our own Galaxy, 
giving a mass around $3\times 10^6\Ms$ \cite{EckhardtGenzel02}.
Here even individual stellar orbits have been measured and shown to revolve
around a single point \cite{EckhardtGenzel02} that coincides with a strong source
of radio and x-ray emission.

In many cases, gas disks have been found orbiting the 
mystery 
object.
For instance, 
H$\alpha$ emission from such a disk in
the galaxy M87 has
revealed a record mass of  $3.2\times 10^9\Ms$ in a region merely 10 light-years across,
and 1.3 cm water maser emission from a disk in the galaxy NGC4258
has revealed $3.6\times 10^7\Ms$ in a region merely 0.42 light-years across
(1 light-year$\>\approx 10^{16}\,$m).
This is too compact to be a stable star cluster, so the only  
alternatives to the black hole explanation involve new physics --- 
like a ``fermion ball'' made of postulated new particles \cite{Munyaneza01}.

Although impressive, all these spacetime measurements were still at
$>10^4$ Schwarzschild radii, and so probe no 
strong GR effects and give only indirect black hole evidence.
X-ray spectroscopy provides another powerful probe, because
x-rays can be produced closer to the event horizon, less than a light hour
from the central engine where the material is hotter and the
detailed shape of spacetime can imprint interesting signatures
on the emitted radiation. 
For instance, a strong emission line from the K$\alpha$ fluorescent
transition of highly (photo-)ionized iron atoms
has been observed 
by the ASCA and  Beppo-SAX satellites 
\cite{Nandra00} 
to have spectacular properties.
Doppler shifts indicate a gas disk rotating with 
velocities up to 10\% of the speed of light, and extremely broadened and 
asymmetric line profiles are best fit when including both
Doppler and gravitational redshifts at 3-10 Schwarzschild radii.

In addition to all this geometric evidence for supermassive black holes,
further support comes from the processes by which they eat and grow.
Infalling gas is predicted to form a hot accretion disk around the hole that 
can radiate away as much as 10\% of its rest energy.
It was indeed this idea that led to the suggestions of supermassive black holes
in the early 1960s, prompted by the discovery of quasars. About
50\% of all galaxies are now known to have active galactic nuclei (AGN)  at least at some low level
--- any black holes in the other half are presumed to have quieted down after consuming the gas in their 
vicinity. 
AGN's can produce luminosities exceeding that of $10^{12}$ suns
in a region less than a light-year across, and
no other mechanism is known for converting matter into radiation
with the high efficiency required.
In some cases, emission has been localized to 
a region $\simlt$ a light-hour across
(smaller than our solar system) by changing intensity in less than an hour.

Furthermore, magnetic phenomena in accretion discs can radiate beams of energetic particles,
and such jets have been observed to up to $10^6$ light-years long, perpendicular
to the disk as predicted.
This requires motions near the speed of light as well as a 
stable preferred axis over long ($\gg 10^6$ year) timescales,
as naturally predicted for black holes \cite{Kormendy01,BlandfordZnajek77}.

\subsection{Stellar-mass black holes}

Numerous stars have been found to orbit a binary companion
weighting too much to be a white dwarf or a neutron star ($\simgt 3\Ms)$,
and being too faint (often invisible) to be a normal star.
For example, after a transient outburst of soft x-rays in 1989, 
all orbital parameters of the binary system V404 Cygni were measured and the black hole
candidate was found to weigh $12\pm 2\Ms$ \cite{Shabhaz94}.
Just as for supermassive BH's, x-ray variability has placed upper limits on
the size of such objects that rule out all conventional black hole alternatives.

To counter such indirect arguments for black holes,
unconventional compact objects such as ``strange stars'' and ``Q-stars''
have been proposed \cite{Witten84,Bahcall90}.
However, the accretion disk model for soft x-ray transients
such as V404 Cygni might require the object to have an event horizon that
gas can disappear through --- a hard surface could
cause radiation to come back out.
Indeed, the similarities between galactic and stellar accretion disk and jet observations
are so striking that a single unified explanation seems natural, 
and black holes provide one.

There is thus strong evidence for existence of black holes in two separate mass ranges, each
making up perhaps $10^{-6}$ or $10^{-5}$ of all mass in the universe. 
Still smaller classes of black holes have been speculated about without direct supporting evidence,
both microscopic ones created in the early universe 
perhaps making up the dark matter \cite{CarrReview} and
transient ones constituting ``spacetime foam'' on the Planck scale \cite{MTW}.

\subsection{Black hole prospects \& gravitational waves }

Whereas it is fairly well-understood how stellar-mass
black holes can be formed by dying massive stars
\cite{OppenheimerSnyder39,Celotti99}, the origin and evolution of the apparently 
ubiquitous supermassive black holes are open questions,
as is their relation to the formation of galaxies and galactic bulges.
Another challenge involves measuring spacetime more accurately near the event horizon,
particularly for evidence of black hole rotation 
\cite{Miller02}.
Observations to look forward to include galactic center flashes as
individual stars get devoured, multiwavelength accretion disk observations,
and, in particular, detection of gravitational waves. 
These tiny ripples in spacetime should be produced whenever 
masses are accelerated, and binary pulsars have been measured to lose energy
at precisely the rate gravitational wave emission predicts \cite{PulsarPaper}.
They should thus be copiously produced in inspiraling mergers involving
black holes, both stellar-mass ones (measurable by ground-based detectors
such as the Laser Interferometer Gravitational wave Observatory, LIGO)
and 
and supermassive ones (measurable by space-based detectors such as 
the Laser Interferometer Space Antenna, LISA) \cite{Hughes01}. 
At still longer wavelengths, the hunt for gravitational waves 
goes on using pulsar timing \cite{Lommen01} and microwave background polarization
that can constrain cosmological inflation \cite{Kamion97,Zalda97}.



\section{Outlook}

I have surveyed recent measurements of spacetime over a factor of $10^{22}$ in
scale, ranging from the cosmic horizon down to the event horizon
of black holes. 
On the largest scales, evidence supports ``back to basics'' flat infinite space
and eternal future time. The growth of spacetime fluctuations has
suggested that about 30\% of the cosmic matter budget is made up of 
(mostly cold) dark matter, about 5\% ordinary matter and the remainder
dark energy. There is further evidence for the same dark matter in 
the halos of galaxies and clusters. Finally, spacetime seems to be full of black holes,
both supermassive ones in the centers of most galaxies and stellar mass ones wherever
high mass stars have died.

How much of this have we really measured and how much is based on assumptions?
The above-mentioned geometric test particle observations have measured the  
spacetime metric, but all inferences about dark energy, dark matter and the 
inner parts of black holes assume that the Einstein Field Equations (EFEs) of GR are valid.
Indeed, attempts have been made to explain away all three by modifying 
the EFEs. So-called scalar-tensor gravity has been found capable of giving accelerated cosmic
expansion without dark energy 
\cite{Boisseau00}. 
Although not an ab initio theory, 
the approach known as Modified Newtonian Dynamics (MOND) attempts to explain galaxy
rotation curves without dark matter 
\cite{Milgrom83,McGaugh00}.
It is not inconceivable that the EFEs can be modified to avoid black hole singularities \cite{Mazur02}, 
even though 
the perhaps most publicized model with this property \cite{Yilmaz} 
has been argued to be flawed \cite{Misner}.

So could  dark energy, dark matter and black 
holes be merely a modern form of epicycles,
which just like
those of Ptolemy can be eliminated by modifying the laws of gravity
\cite{gravity,McGaugh00,Peebles99,Sellwood00}?
The way to answer this question is clearly to test the EFEs observationally, by embedding them in 
a larger class of equations and quantifying the observational constraints.
This program has been pioneered by Clifford Will and others \cite{WillBook,Will98},
showing that the true theory of gravity must be extremely close to GR in the regime probed
by solar system dynamics and binary pulsars, and has also been pursued to 
close the MOND-loophole with some success 
\cite{Griffiths01,WhiteScottPierpaoli01,WhiteKochanek01}. 
However, this does
{\it not} imply that the true theory of gravity must be indistinguishable
from GR in all contexts, in particular for very compact objects \cite{Hughes01} 
or for
cosmology \cite{WillBook,Will98}, so testing gravity remains a fruitful area of research.
Such tests continue even in the laboratory \cite{WashingtonExp}, testing 
the gravitational inverse square law down to millimeter scales to probe possible
extra dimensions \cite{Randall02}.

In conclusion, the coming decade will be exciting:
an avalanche of astrophysical observations are measuring spacetime with unprecedented accuracy,
allowing us to test whether it obeys Einstein's field equations, and consequently whether 
dark energy, dark matter and black holes are for real.

\bigskip
%
%





\section{References and Notes}


\begin{references}
\bibitem{LyAF}
The Lyman $\alpha$ forest  (Ly$\alpha$F) is the plethora of absorption lines in the
spectra of distant quasars caused by neutral hydrogen in overdense
intergalactic gas along the line of sight. It allows us to map
the cosmic gas distribution out to great distances.
\bibitem{powerspectrum}
The CMB power spectrum is the level of temperature fluctuations as a function of angular scale,
defined as the variance of the spherical harmonic coefficients of a sky map.
A matter power spectrum is the level of 3D density fluctuations as a function of spatial scale, defined as the
variance of the Fourier coefficients of the density field.
\bibitem{bias}
The measured power spectrum of galaxies or LyAF clouds may differ from the power
spectrum that we really care about --- that for the underlying dark matter. 
Bias is defined as the square root of the ratio of the two power spectra.
\end{references}
\end{document}